# Diffraction of Coherent Light with Sinusoidal Amplitude by a Thin-Slit Grid

Robert Buonpastore, Ernst Knoesel, and Eduardo Flores
Department of Physics & Astronomy
Rowan University

**Abstract**

We report on the experimental findings of a thin-slit grid diffraction of a coherent light beam with sinusoidal amplitude. We create the sinusoidal amplitude by placing a thin-slit grid at the center of the dark fringes of an interference pattern resulting from laser light passing through a dual pinhole. The experimental results show good agreement with diffraction calculations of a thin-slit grid placed at the center of the dark fringes produced by two coherent beams that intersect at small angle. The results are applied to wire-grid diffraction in the Afshar experiment.

## 1. Introduction

Diffraction calculation of coherent light by a thin-slit is a classical result in optics [1]. However, this calculation has only been tested for cases where the electric field amplitude at the diffracting object is constant. A recent experiment known as the Afshar experiment has demonstrated the need to test a diffraction calculation in which the electric field amplitude varies sinusoidally [2,3]. This controversial experiment that examines the validity of complementarity is the motivation for our work.

The Afshar experiment consists of coherent light incident onto a pair of pinholes [3]. The two beams emerging from the pinholes spatially overlap in the far-field and interfere to produce a pattern of alternating light and dark fringes. At an appropriate distance from the pinholes, thin wires are placed at the minima of the interference pattern. Beyond the wires there is a lens that forms the image of the pinholes onto two photon detectors located at the image of each pinhole.

Wire grid diffraction is minimal in the Afshar experiment, yet it needs to be quantified to obtain the degree of which-way information and estimate the visibility. In this work, we will show that wire grid diffraction can be easily obtained from slit grid diffraction by an application of Babinet's principle. Thus, we will only consider the case of diffraction produced by a thin-slit grid located at a minimum of an interference pattern. For the experiment we use a grid that consists of four thin slits inserted at the minima of the diffraction pattern in the Afshar experiment set-up.

The calculation results have been reported by Flores [4]. Flores used a set up known as the modified Afshar experiment [5] instead of the original Afshar experimental set up. The modified Afshar experiment is a simpler and more transparent version of the Afshar experiment for calculation and analysis purposes. A laser beam impinges on a 50:50 beam splitter and produces two spatially separated coherent beams of equal intensity. The beams overlap at some distance. Beyond the region of overlap the two beams fully separate again (Figure 1). There, two detectors are positioned such that detector 1 detects only photons originating from the mirror, and detector 2 detects only photons originating from the beam splitter. Where the beams overlap they interfere forming a pattern of bright and dark fringes. At the center of the dark fringes we place the thin-slit grid.

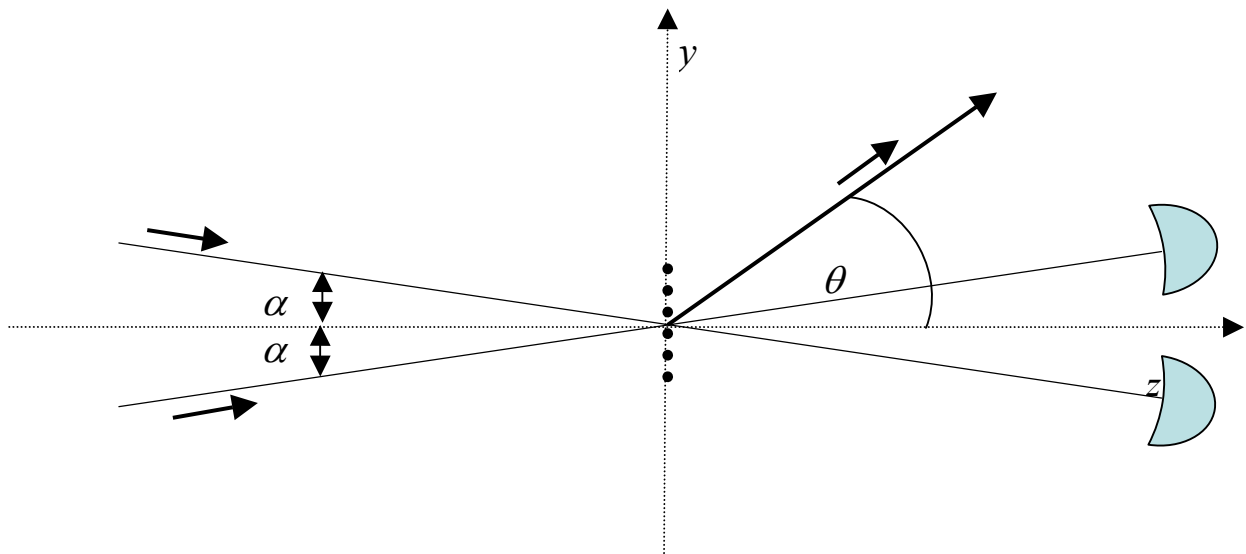

**Figure 1.** In the Modified Afshar Experiment beam separation after interference is achieved by geometry alone without the utilization of a focusing lens as in the Afshar experiment.

Another key aspect of this paper is testing the hypothesis that the Modified Afshar experiment is equivalent to the Afshar experiment. In case of the wire grids, Flores [4] has demonstrated that calculations based on the Modified Afshar experiment agree well with the wire-grid results obtained by Afshar et al. [3] using the original set-up. In this paper we set out to demonstrate that the experimental diffraction pattern for a slit grid using the original Afshar experiment can in fact be reproduced by calculations using the Modified Afshar set-up.

## 2. Summary of the calculation

For the sake of completion we include a summary of the calculation [4]. In the modified Afshar experiment two coherent laser beams propagate on the *y-z* plane and cross each other at the origin (Figure 1). The two plane waves are polarized along the *x*-axis. Throughout this paper we assume a single polarization for the light. Each beam makes an angle $\alpha$ with the z-axis. A slit grid is centered at the origin. The long side of the slits is more than a centimeter long and is parallel to the *x*-axis. Since the width of the beam is less than 5 mm no diffraction takes place along the *x*-axis. Diffraction takes place on the *y-z* plane.

We approximate the two beams as two plane waves with wave vectors $\vec{\kappa}_1 = -\hat{y}\kappa\sin\alpha + \hat{z}\kappa\cos\alpha$ and $\vec{\kappa}_2 = \hat{y}\kappa\sin\alpha + \hat{z}\kappa\cos\alpha$. The superposition of the two plane waves results in a plane wave $E_{eff} \exp(i(\vec{\kappa}'\cdot\vec{r} - \omega t))$, where the wave vector $\vec{\kappa}' = \hat{z}\kappa\cos\alpha$ is directed perpendicular to the slit grid, and $E_{eff}$ is an effective amplitude given by $E_{eff} = 2E_0 \cos(\kappa y \sin\alpha)$. This amplitude shows interference fringes along the y-axis. Near the center of a dark fringe the effective amplitude is approximated by $E_{eff} = \pm 2E_0 \kappa \sin(\alpha) y$, where the sign alternates from slit to slit.

We first calculate the diffraction produced by a single thin slit grid located at the center of a dark fringe of the interference pattern. The classical diffraction integral in the Fraunhofer approximation is [1]

$$\int_{-b/2}^{b/2} \frac{E_{eff}}{r} \sin(\omega t - \kappa r) dy, \qquad (1)$$

where b is the thickness of the slit and $r$ is the distance from the source point to the detector. $r$ is given by $r = (R^2 + y^2 - 2Ry\sin(\theta))^{1/2}$, where $R$ is the distance from the origin to the detector and $\theta$ is the angle that diffracted light makes with the z-axis. Since the phase is much more sensitive to small changes than the amplitude we may replace $r$ in the amplitude by $R$ but $r$ in the phase by $R - y\sin(\theta)$. Thus, the integral is now

$$\Lambda \int_{-b/2}^{b/2} y \sin(\omega t - \kappa R + y\kappa \sin(\theta)) dy, \qquad (2)$$

where $\Lambda = 2E_0 \kappa \sin\alpha / R$. Evaluating this integral gives the electric field at the detector region as function of $\theta$.

The calculation of the electric field produced by N slits, a relatively simple extension of the integral in equation (2), is given by

$$\sum_{j=0}^{j=N-1} (-1)^j \Lambda \int_{jd-b/2}^{jd+b/2} (y - jd) \sin(\omega t - \kappa R + y\kappa \sin(\theta)) dy, \qquad (3)$$

where $d$ is the center-to-center distance between adjacent slits. For our calculation we are more interested in the intensity. We calculate the intensity $I_S(\theta)$ for the $N = 4$ case by taking the time average of the square of the electric field,

$$I_S(\theta) = \frac{2\Lambda^2}{(\kappa \sin\theta)^4} \{b\kappa \sin\theta \cos(1/2 b\kappa \sin\theta) - 2\sin(1/2 b\kappa \sin\theta)\}^2 \times$$
$$\{\sin(1/2 d\kappa \sin\theta) - \sin(3/2 d\kappa \sin\theta)\}^2. \qquad (4)$$

A plot of the intensity $I_S(\theta)$ is presented in Figure 2.

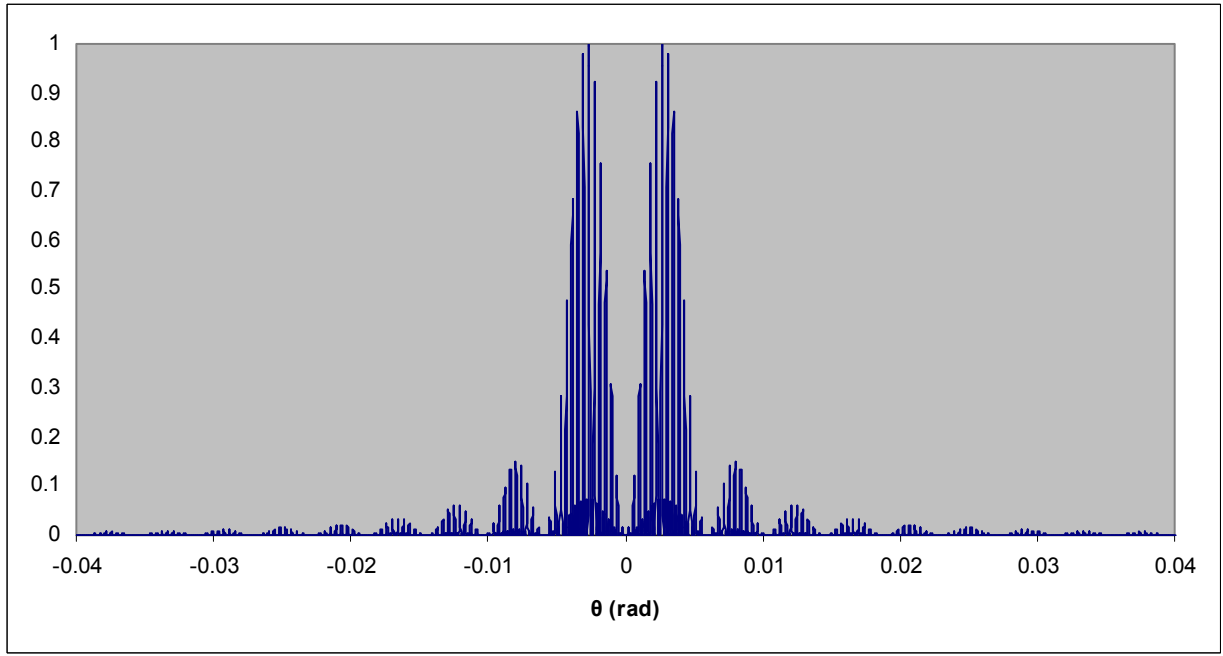

**Figure 2.** Calculated intensity distribution $I_S(\theta)$ for a sinusoidal modulated electric field distribution on a 4-slit grid aligned at the electric field minima.

## 3. Experiment

In our experiment we use green laser light of wavelength $\approx 532$ nm. The beam hits an opaque screen with a pair of pinholes with diameters of 40 μm and center-to-center separation of 250 μm. With the help of an aperture only the first Airy disk is allowed through. A 4-slit grid, each slit 127 μm thick, and center to center separation of 1.3 mm, is located approximately 61 cm from the pinholes. We aligned the grid such that each slit was placed on a dark fringe of the interference pattern. We used lenses to image the resulting diffraction pattern onto an Andor Newton CCD camera in the far field.

We achieved spectroscopic measurements with an exposure time of 200 ms and average over 10 runs. The width of the CCD detector is too small to fully surround the resulting diffraction pattern, so the detector is moved in increments of 1 inch from -2 to +2 inches. We took great care to illuminate both pinholes with equal intensity.

In Figure 3 we present a series of five different data sets pieced together to allow for a full qualitative understanding of the diffracted light. The intensity distribution is far above the stray light levels and highly symmetric. Without the slit grid, the pinholes would be imaged at the very center of the diffraction pattern as the two local center maxima. It is obvious that most of the light is scattered away from the position of the pinholes. The envelope of the pattern resembles a single-slit diffraction pattern but with a pronounced minimum at the center location. This center diffraction minimum is a result of the variable electric field across the slits. The diffraction pattern is superimposed with an interference pattern, which resembles the interference pattern observed for multiple slits. This becomes apparent by the presence of three local interference minima and two local maxima between each two main interference maxima. The near perfect symmetry and the high contrast for the minima leads us to believe that the experiment was correctly aligned with equal intensity on both pinholes.

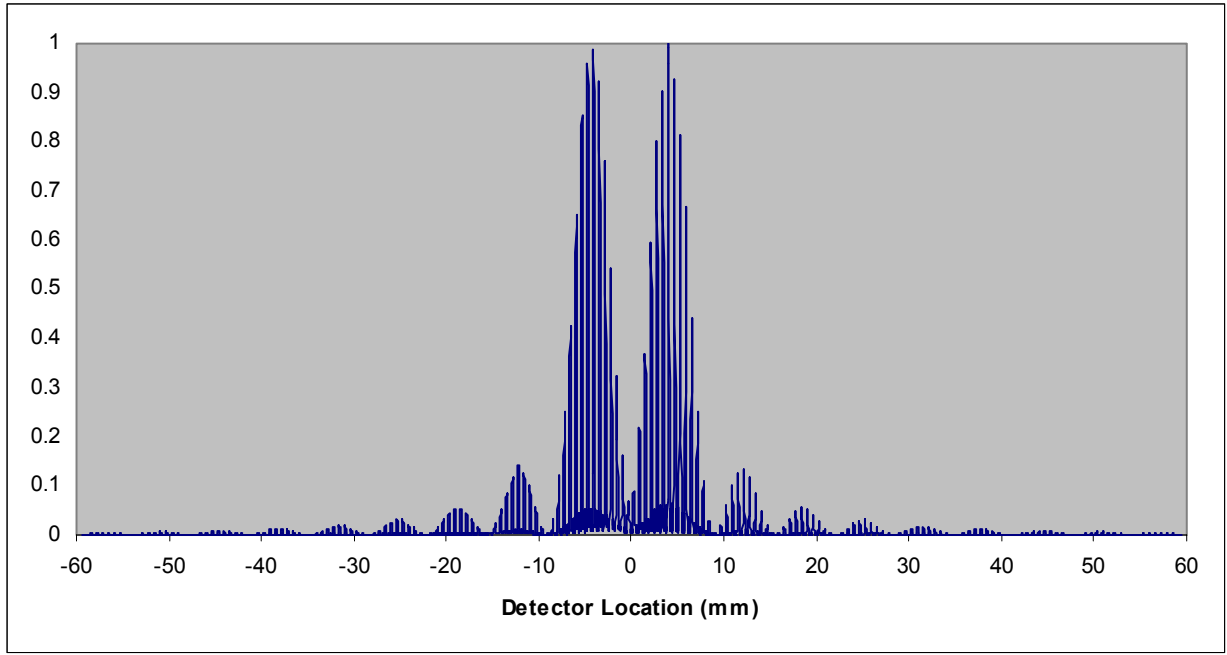

**Figure 3.** Experimental intensity distribution of a 4-slit diffraction pattern with the slit grid positioned at the dark fringes of an interference pattern.

Figure 4 shows a comparison between the theoretical calculations (top) and the experimental results (bottom). The similarities between the two graphs are almost perfect. The main discrepancy can be observed for the two center maximum at the position of the pinholes, where the experimental values are higher than the theoretical calculations. This might be due to the fact that this local maximum is created by destructive interference of large electric fields from the two beams of the pinholes.

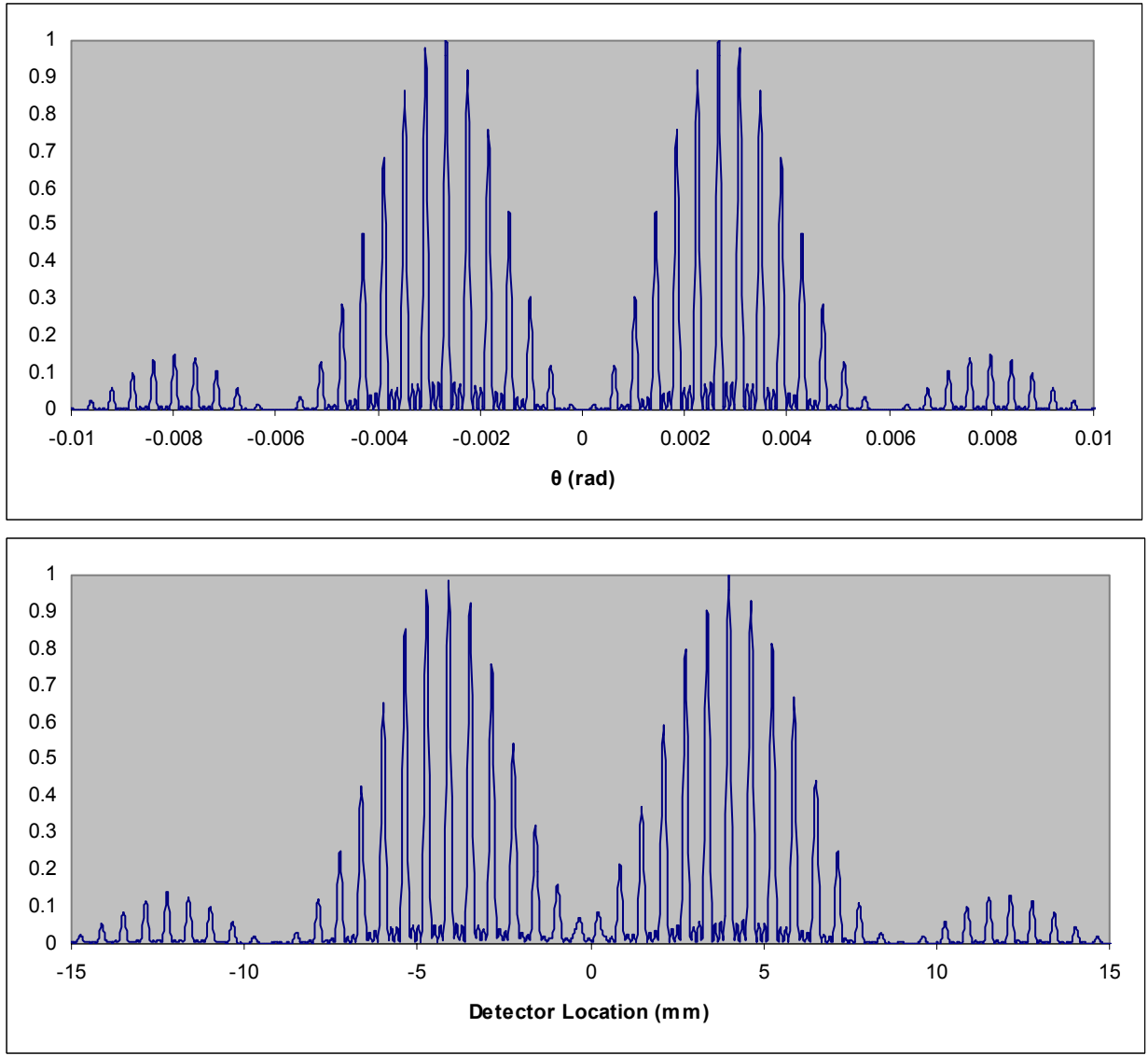

**Figure 4.** Comparison of theory (top) and experiment (bottom) of a 4-slit, varying electric field diffraction pattern. In the absence of the slit grid the unobstructed pinholes would be imaged at the very center of the pattern as the innermost two local maxima.

## 4. Conclusions

The experimental and theoretical results for diffration of a coherent light beam with sinusoidal amplitude by a thin-slit grid are in quite good agreement. Since the theoretical calculation was done using the classical diffraction integral in the Fraunhofer approximation [1], this work extends the validity of this approach to cases where the electric field at the diffracting object is not uniform.

We may apply our results for the thin-slit grid to obtain wire diffraction in the Afshar experiment by invoking Babinet's principle. We call $\vec{E}_0$ the electric field that arrives at the detection region when the laser beam is unperturbed, in other words in the absence of the wire grid or slit grid. Let then $\vec{E}_W$ be the electric field, when the wire-grid is in place. Correspondingly, let $\vec{E}_S$ be the electric field when the complementary slit grid is in place. Then, Babinet's principle states that $\vec{E}_0 = \vec{E}_S + \vec{E}_W$. In our case we are predominantly interested in the region outside the detectors where $\vec{E}_0 = 0$. In this region we get $\vec{E}_W = -\vec{E}_S$, and therefore, the corresponding intensities produced by the wire-grid and the slit grid are identical, $I_W(\theta) = I_S(\theta)$. The function $I_W(\theta)$, and especially the ratio of scattered light off the detectors compared to scattered light on the detectors, has been very instrumental in calculating the degree of which-way information in the Afshar experiment [4].

The excellent agreement of our experimental measurements with the theoretical calculations is a strong argument for the equivalence between the Afshar experiment and the modified Afshar experiment. Previously this equivalence was only tested by the results at the small fixed detectors right in front of the beams at the imaging plane. Now the equivalence test is completed by the check that the interference pattern in the larger region outside the detectors is similar in both set ups. This equivalence test justifies the use of the modified Afshar experiment set up for calculational purposes.